\documentclass[pra,aps,reprint,a4paper,superscriptaddress,floatfix]{revtex4-2}
\usepackage{times}
\usepackage{graphicx}
\usepackage{epsfig}
\usepackage{amsfonts}
\usepackage{amsmath}
\usepackage{amssymb}
\usepackage{dsfont}
\usepackage{amsthm}
\usepackage{color}
\usepackage[colorlinks=true,linkcolor=blue,citecolor=blue,urlcolor=blue]{hyperref}
\usepackage{comment}
\usepackage{braket}
\usepackage{physics}

\begin{document}


\title{Quantum nonlocality: How does nature do it?}
\thanks{Intentionally, the title is the same as that of Ref.\ [28].}


\author{Ad\'an Cabello}
\affiliation{
Departamento de F\'{\i}sica Aplicada II,
Universidad de Sevilla,
41012 Sevilla,
Spain
}
\affiliation{
Instituto Carlos~I de F\'{\i}sica Te\'orica y Computacional, 
Universidad de Sevilla, 
41012 Sevilla, 
Spain
}


\begin{abstract}
In a recent note, Hance and Hossenfelder (arXiv:2211.01331) recall that ``locally causal completions of quantum mechanics are possible, if they violate the assumption [called statistical independence or measurement independence] that the hidden variables do not in any way depend on measurement settings'' and that, consequently, the experimental violations of Bell inequalities ``show that maintaining local causality requires violating statistical independence''. However, Hance and Hossenfelder also argue that ``we should (\ldots) look for independent experimental evidence that can distinguish the two different options: non-locality and statistical independence, or locality and violations of statistical independence'' and that ``the unwillingness to consider theories without statistical independence may be the reason we do not yet have a locally causal theory for the foundations of physics that is consistent with general relativity''.
Here, we recall that there is a third option, namely, rejecting that measurement outcomes are governed in any way by hidden variables. Moreover, we argue that some recent results in the search for principles singling out the sets of quantum correlations for Bell and Kochen-Specker contextuality scenarios point out that this third option is scientifically more plausible and answers the question of why and how nature produces quantum nonlocality.
\end{abstract}



\maketitle


We know that nature violates Bell inequalities \cite{Bell:1964PHY,Clauser:1969PRL,Freedman:1972PRL,Aspect:1982PRL,Weihs:1998PRL,Rowe:2001NAT,Giustina:2013NAT,Hensen:2015NAT,Giustina:2015PRL,Shalm:2015PRL,Rosenfeld:2017PRL},
but we do not know why this violation occurs and what this violation means. There are several possibilities:

(i) It could indicate that measurement outcomes are governed by nonlocal hidden variables \footnote{Here, by nonlocal hidden variables I mean variables that may depend on other spacelike separated variables.} and ``spooky actions at a distance'' \cite{Born:1969XXX}, as proposed in, e.g., \cite{DeBroglie:1925,Bohm:1952PR}.


(ii) It could indicate that measurement outcomes are governed by local hidden variables and measurements depend on the hidden variables, as proposed in, e.g., \cite{Brans:1988IJTP}.

(iii) It could indicate that neither measurements nor measurement outcomes are governed in any way by hidden variables (local or nonlocal).

Here, I propose to examine these possibilities under the light of some recent developments in the program of deriving from fundamental principles the sets of quantum correlations for Bell \cite{Pawlowski:2009NAT,Navascues:2010PRSA,Fritz2013,Navascues:2015NC,Cabello:2019PRA} and Kochen-Specker (KS) \cite{Kochen:1967JMM} contextuality scenarios \cite{CSWPRL2014,Cabello:2013PRL,Cabello:2019PRA}. 

This program aims at finding principles singling out the sets of quantum correlations and, e.g., explain why the maximum violation of the Clauser-Horne-Shimony-Holt (CHSH) Bell inequality \cite{Clauser:1969PRL} (conveniently written) is $2 \sqrt{2} \approx 2.82843$ (as it is apparently the case in nature \cite{Poh:2015PRL}), rather than, e.g., $2.82537$ (as it should be according to some models \cite{Grinbaum:2015FP}) or $4$ (as it could be if the only limiting factor was the principle of no-signaling \cite{Popescu:1994FPH}). A particular branch of this program \cite{CSWPRL2014,Cabello:2013PRL,Cabello:2019PRA}, formulates these questions within a more general framework and relates the question of how nature does Bell nonlocality \cite{Popescu:1994FPH,Gisin:2009Sci} to the question of how nature does KS contextuality \cite{CSWPRL2014}, which also aims at, e.g., explaining why the maximum violation of the Klyachko-Can-Binicio\u{g}lu-Shumovsky (KCBS) noncontextuality inequality \cite{Klyachko:2008PRL} with ideal measurements (and conveniently written) is $\sqrt{5}$. We want to understand exactly why the {\em only} correlations (aka behaviors \cite{Tsirelson:1993HJS} or probability models \cite{Abramsky:2012PRA}) produced by nature for any conceivable Bell or KS contextuality scenario are exactly those allowed by quantum mechanics.

Some of these works have identified very simple general principles which, in the case of ideal measurements of ideal observables \cite{Cabello:2019PRA} \footnote{An ideal (or sharp) measurement of an observable is one which does not disturb any jointly measurable observable (thus it is ``minimally disturbing'' \cite{Chiribella:2016IC}).
An ideal observable is one in which all coarse-grainings can be ideally measured. 
It is important to stress that, independently of how difficult may it be to implement ideal measurements in practice \cite{Guryanova:2020Q} (and, apparently, we can do it pretty well \cite{Pokorny:PRL20,WangSAdv2022}), the concept of ideal observable is necessary in any maximally predictive physical theory allowing joint measurements}, single out all these sets and, in the case of Bell scenarios and treating measurements as black boxes, have, so far, excluded most part of the nonquantum region for some interesting scenarios \cite{Pawlowski:2009NAT,Pawlowski:2009NAT,Navascues:2010PRSA,Fritz2013,Navascues:2015NC}. 

Why these results may be relevant to decide what is the ``message'' \cite{Zeilinger:2005Nat} that quantum mechanics is trying to tell us? First, recall that there is one set of correlations for each measurement scenario \footnote{A measurement scenario is characterized by a number of observables, the cardinality of their respective sets of possible outcomes, and the description of which observables can be jointly measured. For example, the scenario with the smallest number of observables in which there can be KS contextuality (and Bell nonlocality) is the one with four dichotomic observables, $A,B,a,b$, such that the following pairs are jointly measurable: $\{A,B\}$, $\{A,b\}$, $\{a,B\}$, $\{a,b\}$.}. Second, it is important to stress that each of these sets of correlations has a highly complex geometry. To the point that ``we do not have a good intuition about what the [simplest!] quantum set actually looks like'' \cite{Goh:2018PRA}.
Finally, these sets are difficult to bound with mathematical conditions \cite{NPA_PRL}. 

However, e.g., Ref.~\cite{Cabello:2019PRA} shows that the sets of correlations produced by ideal measurements of ideal observables in Bell and KS contextuality scenarios, under the ``principle'' that nature can produce statistically independent copies of every behavior, are exactly those allowed by quantum mechanics. 

Isn't it surprising that we can single out these quantum sets {\em without adding any physical law governing the dynamics of the hidden variables or detailing how measurement independence must be relaxed}?

At the beginning of our search for principles, we might have expected these ``principles'' to encode clues about the type and dynamics of the hidden variables or mechanisms of relaxation of measurement independence. However, the truth is that we have not found anything like this. In particular, we have observed that, in the case of ideal measurements of ideal observables, a universe without hidden variables and with independent experiments {\em would produce exactly the same sets of correlations predicted by quantum mechanics!}

In contrast, universes in which hidden variables (deterministic or stochastic) play a role in the generation of measurement outcomes and in which measurement independence is not strict, would require a {\em different} fine-tuning for producing different points in the different sets of quantum correlations. For example, to obtain $2\sqrt{2}$ for CHSH with local hidden variables, we must relax in a very precise way the assumption of measurement independence \cite{Hall:2010PRL}. However, this relaxation is not the same needed to obtain $\sqrt{5}$ for KCBS or for other points in the set of quantum KS contextual correlations \cite{Hall:2011PRA}.
Causal mechanisms proposed to simulate bipartite Bell nonlocality, including superluminal causal influences (e.g., \cite{DeBroglie:1925,Bohm:1952PR}), relaxations of measurement independence (e.g., \cite{Brans:1988IJTP}), and retrocausality (e.g., \cite{Costa:1977NC,Cramer:1980PRD}), {\em require} fine-tuning \cite{Wood:2015NJP}.

Studying how much superluminal communication must be allowed \cite{Maudlin:1992} or how much measurement independence should be relaxed \cite{Hall:2010PRL} in order to simulate certain correlations is important for, e.g., cryptography, since it allows us to quantify the security against specific attacks. However, if the purpose is answering ``how does nature do it?'' \cite{Gisin:2009Sci}, or ``how come the quantum?'' \cite{Wheeler:1986XXX}, or what is ``the message of the quantum'' \cite{Zeilinger:2005Nat}?, then it seems to be a lesson in the prodigious, miraculous coincidence that, at least at the level of ideal measurements of ideal observables, our universe allows only those correlations that would be natural only if option (iii) happens.

Each of the other possible alternatives carry a heavy ontological baggage that, at least for explaining quantum Bell nonlocality and KS contextuality for ideal measurements of ideal observables, is unnecessary and clashes with useful concepts such as Lorentz invariance and thermodynamics \cite{Cabello_Th_PRA2016}. 
In light of all these arguments, Occam's razor principle strongly suggests that quantum mechanics (and, more precisely, the quantum theory of probabilities) does not need any ``completion''. 

John Wheeler was convinced that ``behind it all is surely an idea so simple, so beautiful, so compelling that when (\ldots) we grasp it, we will all say to each other, how could it have been otherwise?'' \cite{Wheeler:1986XXX}.
The absence of hidden variables behind measurement outcomes is a simple idea. Admittedly, at first sight, option (iii) has little explanatory power. How is it possible that the {\em absence} of something is what explains the highly complex geometry of infinitely many sets? However, this simple idea, properly treated, is the one that actually explains more things. Progress in physics (e.g., a better theory of gravity) requires taking option (iii) into account.


I thank Mateus Ara\'ujo, Michael J. W. Hall, and Emmanuel Zambrini Cruzeiro for comments on earlier versions of this note.


\bibliographystyle{apsrev4-2}
\bibliography{third}

\begin{thebibliography}{48}%
\makeatletter
\providecommand \@ifxundefined [1]{%
 \@ifx{#1\undefined}
}%
\providecommand \@ifnum [1]{%
 \ifnum #1\expandafter \@firstoftwo
 \else \expandafter \@secondoftwo
 \fi
}%
\providecommand \@ifx [1]{%
 \ifx #1\expandafter \@firstoftwo
 \else \expandafter \@secondoftwo
 \fi
}%
\providecommand \natexlab [1]{#1}%
\providecommand \enquote  [1]{``#1''}%
\providecommand \bibnamefont  [1]{#1}%
\providecommand \bibfnamefont [1]{#1}%
\providecommand \citenamefont [1]{#1}%
\providecommand \href@noop [0]{\@secondoftwo}%
\providecommand \href [0]{\begingroup \@sanitize@url \@href}%
\providecommand \@href[1]{\@@startlink{#1}\@@href}%
\providecommand \@@href[1]{\endgroup#1\@@endlink}%
\providecommand \@sanitize@url [0]{\catcode `\\12\catcode `\$12\catcode
  `\&12\catcode `\#12\catcode `\^12\catcode `\_12\catcode `\%12\relax}%
\providecommand \@@startlink[1]{}%
\providecommand \@@endlink[0]{}%
\providecommand \url  [0]{\begingroup\@sanitize@url \@url }%
\providecommand \@url [1]{\endgroup\@href {#1}{\urlprefix }}%
\providecommand \urlprefix  [0]{URL }%
\providecommand \Eprint [0]{\href }%
\providecommand \doibase [0]{https://doi.org/}%
\providecommand \selectlanguage [0]{\@gobble}%
\providecommand \bibinfo  [0]{\@secondoftwo}%
\providecommand \bibfield  [0]{\@secondoftwo}%
\providecommand \translation [1]{[#1]}%
\providecommand \BibitemOpen [0]{}%
\providecommand \bibitemStop [0]{}%
\providecommand \bibitemNoStop [0]{.\EOS\space}%
\providecommand \EOS [0]{\spacefactor3000\relax}%
\providecommand \BibitemShut  [1]{\csname bibitem#1\endcsname}%
\let\auto@bib@innerbib\@empty
\bibitem [{\citenamefont {Bell}(1964)}]{Bell:1964PHY}%
  \BibitemOpen
  \bibfield  {author} {\bibinfo {author} {\bibfnamefont {J.~S.}\ \bibnamefont
  {Bell}},\ }\href {https://doi.org/10.1103/PhysicsPhysiqueFizika.1.195}
  {\bibfield  {journal} {\bibinfo  {journal} {Physics}\ }\textbf {\bibinfo
  {volume} {1}},\ \bibinfo {pages} {195} (\bibinfo {year} {1964})}\BibitemShut
  {NoStop}%
\bibitem [{\citenamefont {Clauser}\ \emph {et~al.}(1969)\citenamefont
  {Clauser}, \citenamefont {Horne}, \citenamefont {Shimony},\ and\
  \citenamefont {Holt}}]{Clauser:1969PRL}%
  \BibitemOpen
  \bibfield  {author} {\bibinfo {author} {\bibfnamefont {J.~F.}\ \bibnamefont
  {Clauser}}, \bibinfo {author} {\bibfnamefont {M.~A.}\ \bibnamefont {Horne}},
  \bibinfo {author} {\bibfnamefont {A.}~\bibnamefont {Shimony}},\ and\ \bibinfo
  {author} {\bibfnamefont {R.~A.}\ \bibnamefont {Holt}},\ }\href
  {https://doi.org/10.1103/PhysRevLett.23.880} {\bibfield  {journal} {\bibinfo
  {journal} {Phys. Rev. Lett.}\ }\textbf {\bibinfo {volume} {23}},\ \bibinfo
  {pages} {880} (\bibinfo {year} {1969})}\BibitemShut {NoStop}%
\bibitem [{\citenamefont {Freedman}\ and\ \citenamefont
  {Clauser}(1972)}]{Freedman:1972PRL}%
  \BibitemOpen
  \bibfield  {author} {\bibinfo {author} {\bibfnamefont {S.~J.}\ \bibnamefont
  {Freedman}}\ and\ \bibinfo {author} {\bibfnamefont {J.~F.}\ \bibnamefont
  {Clauser}},\ }\href {https://doi.org/10.1103/PhysRevLett.28.938} {\bibfield
  {journal} {\bibinfo  {journal} {Phys. Rev. Lett.}\ }\textbf {\bibinfo
  {volume} {28}},\ \bibinfo {pages} {938} (\bibinfo {year} {1972})}\BibitemShut
  {NoStop}%
\bibitem [{\citenamefont {Aspect}\ \emph {et~al.}(1982)\citenamefont {Aspect},
  \citenamefont {Dalibard},\ and\ \citenamefont {Roger}}]{Aspect:1982PRL}%
  \BibitemOpen
  \bibfield  {author} {\bibinfo {author} {\bibfnamefont {A.}~\bibnamefont
  {Aspect}}, \bibinfo {author} {\bibfnamefont {J.}~\bibnamefont {Dalibard}},\
  and\ \bibinfo {author} {\bibfnamefont {G.}~\bibnamefont {Roger}},\ }\href
  {https://doi.org/10.1103/PhysRevLett.49.1804} {\bibfield  {journal} {\bibinfo
   {journal} {Phys. Rev. Lett.}\ }\textbf {\bibinfo {volume} {49}},\ \bibinfo
  {pages} {1804} (\bibinfo {year} {1982})}\BibitemShut {NoStop}%
\bibitem [{\citenamefont {Weihs}\ \emph {et~al.}(1998)\citenamefont {Weihs},
  \citenamefont {Jennewein}, \citenamefont {Simon}, \citenamefont
  {Weinfurter},\ and\ \citenamefont {Zeilinger}}]{Weihs:1998PRL}%
  \BibitemOpen
  \bibfield  {author} {\bibinfo {author} {\bibfnamefont {G.}~\bibnamefont
  {Weihs}}, \bibinfo {author} {\bibfnamefont {T.}~\bibnamefont {Jennewein}},
  \bibinfo {author} {\bibfnamefont {C.}~\bibnamefont {Simon}}, \bibinfo
  {author} {\bibfnamefont {H.}~\bibnamefont {Weinfurter}},\ and\ \bibinfo
  {author} {\bibfnamefont {A.}~\bibnamefont {Zeilinger}},\ }\href
  {https://doi.org/10.1103/PhysRevLett.81.5039} {\bibfield  {journal} {\bibinfo
   {journal} {Phys. Rev. Lett.}\ }\textbf {\bibinfo {volume} {81}},\ \bibinfo
  {pages} {5039} (\bibinfo {year} {1998})}\BibitemShut {NoStop}%
\bibitem [{\citenamefont {Rowe}\ \emph {et~al.}(2001)\citenamefont {Rowe},
  \citenamefont {Kielpinski}, \citenamefont {Meyer}, \citenamefont {Sackett},
  \citenamefont {Itano}, \citenamefont {Monroe},\ and\ \citenamefont
  {Wineland}}]{Rowe:2001NAT}%
  \BibitemOpen
  \bibfield  {author} {\bibinfo {author} {\bibfnamefont {M.~A.}\ \bibnamefont
  {Rowe}}, \bibinfo {author} {\bibfnamefont {D.}~\bibnamefont {Kielpinski}},
  \bibinfo {author} {\bibfnamefont {V.}~\bibnamefont {Meyer}}, \bibinfo
  {author} {\bibfnamefont {C.~A.}\ \bibnamefont {Sackett}}, \bibinfo {author}
  {\bibfnamefont {W.~M.}\ \bibnamefont {Itano}}, \bibinfo {author}
  {\bibfnamefont {C.}~\bibnamefont {Monroe}},\ and\ \bibinfo {author}
  {\bibfnamefont {D.~J.}\ \bibnamefont {Wineland}},\ }\href
  {https://doi.org/10.1038/35057215} {\bibfield  {journal} {\bibinfo  {journal}
  {Nature}\ }\textbf {\bibinfo {volume} {409}},\ \bibinfo {pages} {791}
  (\bibinfo {year} {2001})}\BibitemShut {NoStop}%
\bibitem [{\citenamefont {Giustina}\ \emph {et~al.}(2013)\citenamefont
  {Giustina}, \citenamefont {Mech}, \citenamefont {Ramelow}, \citenamefont
  {Wittmann}, \citenamefont {Kofler}, \citenamefont {Beyer}, \citenamefont
  {Lita}, \citenamefont {Calkins}, \citenamefont {Gerrits}, \citenamefont
  {Nam}, \citenamefont {Ursin},\ and\ \citenamefont
  {Zeilinger}}]{Giustina:2013NAT}%
  \BibitemOpen
  \bibfield  {author} {\bibinfo {author} {\bibfnamefont {M.}~\bibnamefont
  {Giustina}}, \bibinfo {author} {\bibfnamefont {A.}~\bibnamefont {Mech}},
  \bibinfo {author} {\bibfnamefont {S.}~\bibnamefont {Ramelow}}, \bibinfo
  {author} {\bibfnamefont {B.}~\bibnamefont {Wittmann}}, \bibinfo {author}
  {\bibfnamefont {J.}~\bibnamefont {Kofler}}, \bibinfo {author} {\bibfnamefont
  {J.}~\bibnamefont {Beyer}}, \bibinfo {author} {\bibfnamefont
  {A.}~\bibnamefont {Lita}}, \bibinfo {author} {\bibfnamefont {B.}~\bibnamefont
  {Calkins}}, \bibinfo {author} {\bibfnamefont {T.}~\bibnamefont {Gerrits}},
  \bibinfo {author} {\bibfnamefont {S.~W.}\ \bibnamefont {Nam}}, \bibinfo
  {author} {\bibfnamefont {R.}~\bibnamefont {Ursin}},\ and\ \bibinfo {author}
  {\bibfnamefont {A.}~\bibnamefont {Zeilinger}},\ }\href
  {https://doi.org/10.1038/nature12012} {\bibfield  {journal} {\bibinfo
  {journal} {Nature}\ }\textbf {\bibinfo {volume} {497}},\ \bibinfo {pages}
  {227} (\bibinfo {year} {2013})}\BibitemShut {NoStop}%
\bibitem [{\citenamefont {Hensen}\ \emph {et~al.}(2015)\citenamefont {Hensen},
  \citenamefont {Bernien}, \citenamefont {Dr\'eau}, \citenamefont {Reiserer},
  \citenamefont {Kalb}, \citenamefont {Blok}, \citenamefont {Ruitenberg},
  \citenamefont {Vermeulen}, \citenamefont {Schouten}, \citenamefont
  {Abell\'an}, \citenamefont {Amaya}, \citenamefont {Pruneri}, \citenamefont
  {Mitchell}, \citenamefont {Markham}, \citenamefont {Twitchen}, \citenamefont
  {Elkouss}, \citenamefont {Wehner}, \citenamefont {Taminiau},\ and\
  \citenamefont {Hanson}}]{Hensen:2015NAT}%
  \BibitemOpen
  \bibfield  {author} {\bibinfo {author} {\bibfnamefont {B.}~\bibnamefont
  {Hensen}}, \bibinfo {author} {\bibfnamefont {H.}~\bibnamefont {Bernien}},
  \bibinfo {author} {\bibfnamefont {A.~E.}\ \bibnamefont {Dr\'eau}}, \bibinfo
  {author} {\bibfnamefont {A.}~\bibnamefont {Reiserer}}, \bibinfo {author}
  {\bibfnamefont {N.}~\bibnamefont {Kalb}}, \bibinfo {author} {\bibfnamefont
  {M.~S.}\ \bibnamefont {Blok}}, \bibinfo {author} {\bibfnamefont
  {J.}~\bibnamefont {Ruitenberg}}, \bibinfo {author} {\bibfnamefont {R.~F.~L.}\
  \bibnamefont {Vermeulen}}, \bibinfo {author} {\bibfnamefont {R.~N.}\
  \bibnamefont {Schouten}}, \bibinfo {author} {\bibfnamefont {C.}~\bibnamefont
  {Abell\'an}}, \bibinfo {author} {\bibfnamefont {W.}~\bibnamefont {Amaya}},
  \bibinfo {author} {\bibfnamefont {V.}~\bibnamefont {Pruneri}}, \bibinfo
  {author} {\bibfnamefont {M.~W.}\ \bibnamefont {Mitchell}}, \bibinfo {author}
  {\bibfnamefont {M.}~\bibnamefont {Markham}}, \bibinfo {author} {\bibfnamefont
  {D.~J.}\ \bibnamefont {Twitchen}}, \bibinfo {author} {\bibfnamefont
  {D.}~\bibnamefont {Elkouss}}, \bibinfo {author} {\bibfnamefont
  {S.}~\bibnamefont {Wehner}}, \bibinfo {author} {\bibfnamefont {T.~H.}\
  \bibnamefont {Taminiau}},\ and\ \bibinfo {author} {\bibfnamefont
  {R.}~\bibnamefont {Hanson}},\ }\href {https://doi.org/10.1038/nature15759}
  {\bibfield  {journal} {\bibinfo  {journal} {Nature}\ }\textbf {\bibinfo
  {volume} {526}},\ \bibinfo {pages} {682} (\bibinfo {year}
  {2015})}\BibitemShut {NoStop}%
\bibitem [{\citenamefont {Giustina}\ \emph {et~al.}(2015)\citenamefont
  {Giustina}, \citenamefont {Versteegh}, \citenamefont {Wengerowsky},
  \citenamefont {Handsteiner}, \citenamefont {Hochrainer}, \citenamefont
  {Phelan}, \citenamefont {Steinlechner}, \citenamefont {Kofler}, \citenamefont
  {Larsson}, \citenamefont {Abell\'an}, \citenamefont {Amaya}, \citenamefont
  {Pruneri}, \citenamefont {Mitchell}, \citenamefont {Beyer}, \citenamefont
  {Gerrits}, \citenamefont {Lita}, \citenamefont {Shalm}, \citenamefont {Nam},
  \citenamefont {Scheidl}, \citenamefont {Ursin}, \citenamefont {Wittmann},\
  and\ \citenamefont {Zeilinger}}]{Giustina:2015PRL}%
  \BibitemOpen
  \bibfield  {author} {\bibinfo {author} {\bibfnamefont {M.}~\bibnamefont
  {Giustina}}, \bibinfo {author} {\bibfnamefont {M.~A.~M.}\ \bibnamefont
  {Versteegh}}, \bibinfo {author} {\bibfnamefont {S.}~\bibnamefont
  {Wengerowsky}}, \bibinfo {author} {\bibfnamefont {J.}~\bibnamefont
  {Handsteiner}}, \bibinfo {author} {\bibfnamefont {A.}~\bibnamefont
  {Hochrainer}}, \bibinfo {author} {\bibfnamefont {K.}~\bibnamefont {Phelan}},
  \bibinfo {author} {\bibfnamefont {F.}~\bibnamefont {Steinlechner}}, \bibinfo
  {author} {\bibfnamefont {J.}~\bibnamefont {Kofler}}, \bibinfo {author}
  {\bibfnamefont {J.-A.}\ \bibnamefont {Larsson}}, \bibinfo {author}
  {\bibfnamefont {C.}~\bibnamefont {Abell\'an}}, \bibinfo {author}
  {\bibfnamefont {W.}~\bibnamefont {Amaya}}, \bibinfo {author} {\bibfnamefont
  {V.}~\bibnamefont {Pruneri}}, \bibinfo {author} {\bibfnamefont {M.~W.}\
  \bibnamefont {Mitchell}}, \bibinfo {author} {\bibfnamefont {J.}~\bibnamefont
  {Beyer}}, \bibinfo {author} {\bibfnamefont {T.}~\bibnamefont {Gerrits}},
  \bibinfo {author} {\bibfnamefont {A.~E.}\ \bibnamefont {Lita}}, \bibinfo
  {author} {\bibfnamefont {L.~K.}\ \bibnamefont {Shalm}}, \bibinfo {author}
  {\bibfnamefont {S.~W.}\ \bibnamefont {Nam}}, \bibinfo {author} {\bibfnamefont
  {T.}~\bibnamefont {Scheidl}}, \bibinfo {author} {\bibfnamefont
  {R.}~\bibnamefont {Ursin}}, \bibinfo {author} {\bibfnamefont
  {B.}~\bibnamefont {Wittmann}},\ and\ \bibinfo {author} {\bibfnamefont
  {A.}~\bibnamefont {Zeilinger}},\ }\href
  {https://doi.org/10.1103/PhysRevLett.115.250401} {\bibfield  {journal}
  {\bibinfo  {journal} {Phys. Rev. Lett.}\ }\textbf {\bibinfo {volume} {115}},\
  \bibinfo {pages} {250401} (\bibinfo {year} {2015})}\BibitemShut {NoStop}%
\bibitem [{\citenamefont {Shalm}\ \emph {et~al.}(2015)\citenamefont {Shalm},
  \citenamefont {Meyer-Scott}, \citenamefont {Christensen}, \citenamefont
  {Bierhorst}, \citenamefont {Wayne}, \citenamefont {Stevens}, \citenamefont
  {Gerrits}, \citenamefont {Glancy}, \citenamefont {Hamel}, \citenamefont
  {Allman}, \citenamefont {Coakley}, \citenamefont {Dyer}, \citenamefont
  {Hodge}, \citenamefont {Lita}, \citenamefont {Verma}, \citenamefont
  {Lambrocco}, \citenamefont {Tortorici}, \citenamefont {Migdall},
  \citenamefont {Zhang}, \citenamefont {Kumor}, \citenamefont {Farr},
  \citenamefont {Marsili}, \citenamefont {Shaw}, \citenamefont {Stern},
  \citenamefont {Abell\'an}, \citenamefont {Amaya}, \citenamefont {Pruneri},
  \citenamefont {Jennewein}, \citenamefont {Mitchell}, \citenamefont {Kwiat},
  \citenamefont {Bienfang}, \citenamefont {Mirin}, \citenamefont {Knill},\ and\
  \citenamefont {Nam}}]{Shalm:2015PRL}%
  \BibitemOpen
  \bibfield  {author} {\bibinfo {author} {\bibfnamefont {L.~K.}\ \bibnamefont
  {Shalm}}, \bibinfo {author} {\bibfnamefont {E.}~\bibnamefont {Meyer-Scott}},
  \bibinfo {author} {\bibfnamefont {B.~G.}\ \bibnamefont {Christensen}},
  \bibinfo {author} {\bibfnamefont {P.}~\bibnamefont {Bierhorst}}, \bibinfo
  {author} {\bibfnamefont {M.~A.}\ \bibnamefont {Wayne}}, \bibinfo {author}
  {\bibfnamefont {M.~J.}\ \bibnamefont {Stevens}}, \bibinfo {author}
  {\bibfnamefont {T.}~\bibnamefont {Gerrits}}, \bibinfo {author} {\bibfnamefont
  {S.}~\bibnamefont {Glancy}}, \bibinfo {author} {\bibfnamefont {D.~R.}\
  \bibnamefont {Hamel}}, \bibinfo {author} {\bibfnamefont {M.~S.}\ \bibnamefont
  {Allman}}, \bibinfo {author} {\bibfnamefont {K.~J.}\ \bibnamefont {Coakley}},
  \bibinfo {author} {\bibfnamefont {S.~D.}\ \bibnamefont {Dyer}}, \bibinfo
  {author} {\bibfnamefont {C.}~\bibnamefont {Hodge}}, \bibinfo {author}
  {\bibfnamefont {A.~E.}\ \bibnamefont {Lita}}, \bibinfo {author}
  {\bibfnamefont {V.~B.}\ \bibnamefont {Verma}}, \bibinfo {author}
  {\bibfnamefont {C.}~\bibnamefont {Lambrocco}}, \bibinfo {author}
  {\bibfnamefont {E.}~\bibnamefont {Tortorici}}, \bibinfo {author}
  {\bibfnamefont {A.~L.}\ \bibnamefont {Migdall}}, \bibinfo {author}
  {\bibfnamefont {Y.}~\bibnamefont {Zhang}}, \bibinfo {author} {\bibfnamefont
  {D.~R.}\ \bibnamefont {Kumor}}, \bibinfo {author} {\bibfnamefont {W.~H.}\
  \bibnamefont {Farr}}, \bibinfo {author} {\bibfnamefont {F.}~\bibnamefont
  {Marsili}}, \bibinfo {author} {\bibfnamefont {M.~D.}\ \bibnamefont {Shaw}},
  \bibinfo {author} {\bibfnamefont {J.~A.}\ \bibnamefont {Stern}}, \bibinfo
  {author} {\bibfnamefont {C.}~\bibnamefont {Abell\'an}}, \bibinfo {author}
  {\bibfnamefont {W.}~\bibnamefont {Amaya}}, \bibinfo {author} {\bibfnamefont
  {V.}~\bibnamefont {Pruneri}}, \bibinfo {author} {\bibfnamefont
  {T.}~\bibnamefont {Jennewein}}, \bibinfo {author} {\bibfnamefont {M.~W.}\
  \bibnamefont {Mitchell}}, \bibinfo {author} {\bibfnamefont {P.~G.}\
  \bibnamefont {Kwiat}}, \bibinfo {author} {\bibfnamefont {J.~C.}\ \bibnamefont
  {Bienfang}}, \bibinfo {author} {\bibfnamefont {R.~P.}\ \bibnamefont {Mirin}},
  \bibinfo {author} {\bibfnamefont {E.}~\bibnamefont {Knill}},\ and\ \bibinfo
  {author} {\bibfnamefont {S.~W.}\ \bibnamefont {Nam}},\ }\href
  {https://doi.org/10.1103/PhysRevLett.115.250402} {\bibfield  {journal}
  {\bibinfo  {journal} {Phys. Rev. Lett.}\ }\textbf {\bibinfo {volume} {115}},\
  \bibinfo {pages} {250402} (\bibinfo {year} {2015})}\BibitemShut {NoStop}%
\bibitem [{\citenamefont {Rosenfeld}\ \emph {et~al.}(2017)\citenamefont
  {Rosenfeld}, \citenamefont {Burchardt}, \citenamefont {Garthoff},
  \citenamefont {Redeker}, \citenamefont {Ortegel}, \citenamefont {Rau},\ and\
  \citenamefont {Weinfurter}}]{Rosenfeld:2017PRL}%
  \BibitemOpen
  \bibfield  {author} {\bibinfo {author} {\bibfnamefont {W.}~\bibnamefont
  {Rosenfeld}}, \bibinfo {author} {\bibfnamefont {D.}~\bibnamefont
  {Burchardt}}, \bibinfo {author} {\bibfnamefont {R.}~\bibnamefont {Garthoff}},
  \bibinfo {author} {\bibfnamefont {K.}~\bibnamefont {Redeker}}, \bibinfo
  {author} {\bibfnamefont {N.}~\bibnamefont {Ortegel}}, \bibinfo {author}
  {\bibfnamefont {M.}~\bibnamefont {Rau}},\ and\ \bibinfo {author}
  {\bibfnamefont {H.}~\bibnamefont {Weinfurter}},\ }\href
  {https://doi.org/10.1103/PhysRevLett.119.010402} {\bibfield  {journal}
  {\bibinfo  {journal} {Phys. Rev. Lett.}\ }\textbf {\bibinfo {volume} {119}},\
  \bibinfo {pages} {010402} (\bibinfo {year} {2017})}\BibitemShut {NoStop}%
\bibitem [{Note1()}]{Note1}%
  \BibitemOpen
  \bibinfo {note} {Here, by nonlocal hidden variables I mean variables that may
  depend on other spacelike separated variables.}\BibitemShut {Stop}%
\bibitem [{\citenamefont {Einstein}(1969)}]{Born:1969XXX}%
  \BibitemOpen
  \bibfield  {author} {\bibinfo {author} {\bibfnamefont {A.}~\bibnamefont
  {Einstein}},\ }in\ \href@noop {} {\emph {\bibinfo {booktitle} {Der
  Einstein-Born Briefwechsel 1916--1955}}},\ \bibinfo {editor} {edited by\
  \bibinfo {editor} {\bibfnamefont {M.}~\bibnamefont {Born}}\ and\ \bibinfo
  {editor} {\bibfnamefont {A.}~\bibnamefont {Einstein}}}\ (\bibinfo
  {publisher} {Nymphenburger},\ \bibinfo {address} {M\"unchen},\ \bibinfo
  {year} {1969})\BibitemShut {NoStop}%
\bibitem [{\citenamefont {de~Broglie}(1925)}]{DeBroglie:1925}%
  \BibitemOpen
  \bibfield  {author} {\bibinfo {author} {\bibfnamefont {L.}~\bibnamefont
  {de~Broglie}},\ }\href {https://doi.org/10.1051/anphys/192510030022}
  {\bibfield  {journal} {\bibinfo  {journal} {Ann. Phys.}\ }\textbf {\bibinfo
  {volume} {10}},\ \bibinfo {pages} {22} (\bibinfo {year} {1925})}\BibitemShut
  {NoStop}%
\bibitem [{\citenamefont {Bohm}(1952)}]{Bohm:1952PR}%
  \BibitemOpen
  \bibfield  {author} {\bibinfo {author} {\bibfnamefont {D.}~\bibnamefont
  {Bohm}},\ }\href {https://doi.org/10.1103/PhysRev.85.166} {\bibfield
  {journal} {\bibinfo  {journal} {Phys. Rev.}\ }\textbf {\bibinfo {volume}
  {85}},\ \bibinfo {pages} {166} (\bibinfo {year} {1952})}\BibitemShut
  {NoStop}%
\bibitem [{\citenamefont {Brans}(1988)}]{Brans:1988IJTP}%
  \BibitemOpen
  \bibfield  {author} {\bibinfo {author} {\bibfnamefont {C.}~\bibnamefont
  {Brans}},\ }\href {https://doi.org/10.1007/BF00670750} {\bibfield  {journal}
  {\bibinfo  {journal} {Int. J. Theor. Phys,}\ }\textbf {\bibinfo {volume}
  {27}},\ \bibinfo {pages} {219} (\bibinfo {year} {1988})}\BibitemShut
  {NoStop}%
\bibitem [{\citenamefont {Paw{\l}owski}\ \emph {et~al.}(2009)\citenamefont
  {Paw{\l}owski}, \citenamefont {Paterek}, \citenamefont {Kaszlikowski},
  \citenamefont {Scarani}, \citenamefont {Winter},\ and\ \citenamefont
  {{\.Z}ukowski}}]{Pawlowski:2009NAT}%
  \BibitemOpen
  \bibfield  {author} {\bibinfo {author} {\bibfnamefont {M.}~\bibnamefont
  {Paw{\l}owski}}, \bibinfo {author} {\bibfnamefont {T.}~\bibnamefont
  {Paterek}}, \bibinfo {author} {\bibfnamefont {D.}~\bibnamefont
  {Kaszlikowski}}, \bibinfo {author} {\bibfnamefont {V.}~\bibnamefont
  {Scarani}}, \bibinfo {author} {\bibfnamefont {A.}~\bibnamefont {Winter}},\
  and\ \bibinfo {author} {\bibfnamefont {M.}~\bibnamefont {{\.Z}ukowski}},\
  }\href {https://doi.org/10.1038/nature08400} {\bibfield  {journal} {\bibinfo
  {journal} {Nature}\ }\textbf {\bibinfo {volume} {461}},\ \bibinfo {pages}
  {1101} (\bibinfo {year} {2009})}\BibitemShut {NoStop}%
\bibitem [{\citenamefont {Navascu{\'e}s}\ and\ \citenamefont
  {Wunderlich}(2010)}]{Navascues:2010PRSA}%
  \BibitemOpen
  \bibfield  {author} {\bibinfo {author} {\bibfnamefont {M.}~\bibnamefont
  {Navascu{\'e}s}}\ and\ \bibinfo {author} {\bibfnamefont {H.}~\bibnamefont
  {Wunderlich}},\ }\href {https://doi.org/10.1098/rspa.2009.0453} {\bibfield
  {journal} {\bibinfo  {journal} {Proc. R. Soc. A}\ }\textbf {\bibinfo {volume}
  {466}},\ \bibinfo {pages} {881} (\bibinfo {year} {2010})}\BibitemShut
  {NoStop}%
\bibitem [{\citenamefont {Fritz}\ \emph {et~al.}(2013)\citenamefont {Fritz},
  \citenamefont {Sainz}, \citenamefont {Augusiak}, \citenamefont {Brask},
  \citenamefont {Chaves}, \citenamefont {Leverrier},\ and\ \citenamefont
  {Ac{\'\i}n}}]{Fritz2013}%
  \BibitemOpen
  \bibfield  {author} {\bibinfo {author} {\bibfnamefont {T.}~\bibnamefont
  {Fritz}}, \bibinfo {author} {\bibfnamefont {A.~B.}\ \bibnamefont {Sainz}},
  \bibinfo {author} {\bibfnamefont {R.}~\bibnamefont {Augusiak}}, \bibinfo
  {author} {\bibfnamefont {J.~B.}\ \bibnamefont {Brask}}, \bibinfo {author}
  {\bibfnamefont {R.}~\bibnamefont {Chaves}}, \bibinfo {author} {\bibfnamefont
  {A.}~\bibnamefont {Leverrier}},\ and\ \bibinfo {author} {\bibfnamefont
  {A.}~\bibnamefont {Ac{\'\i}n}},\ }\href {https://doi.org/10.1038/ncomms3263}
  {\bibfield  {journal} {\bibinfo  {journal} {Nat. Commun.}\ }\textbf {\bibinfo
  {volume} {4}},\ \bibinfo {pages} {1} (\bibinfo {year} {2013})}\BibitemShut
  {NoStop}%
\bibitem [{\citenamefont {Navascu{\'e}s}\ \emph {et~al.}(2015)\citenamefont
  {Navascu{\'e}s}, \citenamefont {Guryanova}, \citenamefont {Hoban},\ and\
  \citenamefont {Ac{\'\i}n}}]{Navascues:2015NC}%
  \BibitemOpen
  \bibfield  {author} {\bibinfo {author} {\bibfnamefont {M.}~\bibnamefont
  {Navascu{\'e}s}}, \bibinfo {author} {\bibfnamefont {Y.}~\bibnamefont
  {Guryanova}}, \bibinfo {author} {\bibfnamefont {M.~J.}\ \bibnamefont
  {Hoban}},\ and\ \bibinfo {author} {\bibfnamefont {A.}~\bibnamefont
  {Ac{\'\i}n}},\ }\href {https://doi.org/10.1038/ncomms7288} {\bibfield
  {journal} {\bibinfo  {journal} {Nat. Commun.}\ }\textbf {\bibinfo {volume}
  {6}},\ \bibinfo {pages} {6288} (\bibinfo {year} {2015})}\BibitemShut
  {NoStop}%
\bibitem [{\citenamefont {Cabello}(2019)}]{Cabello:2019PRA}%
  \BibitemOpen
  \bibfield  {author} {\bibinfo {author} {\bibfnamefont {A.}~\bibnamefont
  {Cabello}},\ }\href {https://doi.org/10.1103/PhysRevA.100.032120} {\bibfield
  {journal} {\bibinfo  {journal} {Phys. Rev. A}\ }\textbf {\bibinfo {volume}
  {100}},\ \bibinfo {pages} {032120} (\bibinfo {year} {2019})}\BibitemShut
  {NoStop}%
\bibitem [{\citenamefont {Kochen}\ and\ \citenamefont
  {Specker}(1967)}]{Kochen:1967JMM}%
  \BibitemOpen
  \bibfield  {author} {\bibinfo {author} {\bibfnamefont {S.}~\bibnamefont
  {Kochen}}\ and\ \bibinfo {author} {\bibfnamefont {E.~P.}\ \bibnamefont
  {Specker}},\ }\href {https://doi.org/10.1512/iumj.1968.17.17004} {\bibfield
  {journal} {\bibinfo  {journal} {J. Math. Mech.}\ }\textbf {\bibinfo {volume}
  {17}},\ \bibinfo {pages} {59} (\bibinfo {year} {1967})}\BibitemShut {NoStop}%
\bibitem [{\citenamefont {Cabello}\ \emph {et~al.}(2014)\citenamefont
  {Cabello}, \citenamefont {Severini},\ and\ \citenamefont
  {Winter}}]{CSWPRL2014}%
  \BibitemOpen
  \bibfield  {author} {\bibinfo {author} {\bibfnamefont {A.}~\bibnamefont
  {Cabello}}, \bibinfo {author} {\bibfnamefont {S.}~\bibnamefont {Severini}},\
  and\ \bibinfo {author} {\bibfnamefont {A.}~\bibnamefont {Winter}},\ }\href
  {https://doi.org/10.1103/PhysRevLett.112.040401} {\bibfield  {journal}
  {\bibinfo  {journal} {Phys. Rev. Lett.}\ }\textbf {\bibinfo {volume} {112}},\
  \bibinfo {pages} {040401} (\bibinfo {year} {2014})}\BibitemShut {NoStop}%
\bibitem [{\citenamefont {Cabello}(2013)}]{Cabello:2013PRL}%
  \BibitemOpen
  \bibfield  {author} {\bibinfo {author} {\bibfnamefont {A.}~\bibnamefont
  {Cabello}},\ }\href {https://doi.org/10.1103/PhysRevLett.110.060402}
  {\bibfield  {journal} {\bibinfo  {journal} {Phys. Rev. Lett.}\ }\textbf
  {\bibinfo {volume} {110}},\ \bibinfo {pages} {060402} (\bibinfo {year}
  {2013})}\BibitemShut {NoStop}%
\bibitem [{\citenamefont {Poh}\ \emph {et~al.}(2015)\citenamefont {Poh},
  \citenamefont {Joshi}, \citenamefont {Cer\`e}, \citenamefont {Cabello},\ and\
  \citenamefont {Kurtsiefer}}]{Poh:2015PRL}%
  \BibitemOpen
  \bibfield  {author} {\bibinfo {author} {\bibfnamefont {H.~S.}\ \bibnamefont
  {Poh}}, \bibinfo {author} {\bibfnamefont {S.~K.}\ \bibnamefont {Joshi}},
  \bibinfo {author} {\bibfnamefont {A.}~\bibnamefont {Cer\`e}}, \bibinfo
  {author} {\bibfnamefont {A.}~\bibnamefont {Cabello}},\ and\ \bibinfo {author}
  {\bibfnamefont {C.}~\bibnamefont {Kurtsiefer}},\ }\href
  {https://doi.org/10.1103/PhysRevLett.115.180408} {\bibfield  {journal}
  {\bibinfo  {journal} {Phys. Rev. Lett.}\ }\textbf {\bibinfo {volume} {115}},\
  \bibinfo {pages} {180408} (\bibinfo {year} {2015})}\BibitemShut {NoStop}%
\bibitem [{\citenamefont {Grinbaum}(2015)}]{Grinbaum:2015FP}%
  \BibitemOpen
  \bibfield  {author} {\bibinfo {author} {\bibfnamefont {A.}~\bibnamefont
  {Grinbaum}},\ }\href {https://doi.org/10.1007/s10701-015-9937-y} {\bibfield
  {journal} {\bibinfo  {journal} {Found. Phys.}\ }\textbf {\bibinfo {volume}
  {45}},\ \bibinfo {pages} {1341} (\bibinfo {year} {2015})}\BibitemShut
  {NoStop}%
\bibitem [{\citenamefont {Popescu}\ and\ \citenamefont
  {Rohrlich}(1994)}]{Popescu:1994FPH}%
  \BibitemOpen
  \bibfield  {author} {\bibinfo {author} {\bibfnamefont {S.}~\bibnamefont
  {Popescu}}\ and\ \bibinfo {author} {\bibfnamefont {D.}~\bibnamefont
  {Rohrlich}},\ }\href {https://doi.org/10.1007/BF02058098} {\bibfield
  {journal} {\bibinfo  {journal} {Found. Phys.}\ }\textbf {\bibinfo {volume}
  {24}},\ \bibinfo {pages} {379} (\bibinfo {year} {1994})}\BibitemShut
  {NoStop}%
\bibitem [{\citenamefont {Gisin}(2009)}]{Gisin:2009Sci}%
  \BibitemOpen
  \bibfield  {author} {\bibinfo {author} {\bibfnamefont {N.}~\bibnamefont
  {Gisin}},\ }\href {https://doi.org/10.1126/science.1182103} {\bibfield
  {journal} {\bibinfo  {journal} {Science}\ }\textbf {\bibinfo {volume}
  {326}},\ \bibinfo {pages} {1357} (\bibinfo {year} {2009})}\BibitemShut
  {NoStop}%
\bibitem [{\citenamefont {Klyachko}\ \emph {et~al.}(2008)\citenamefont
  {Klyachko}, \citenamefont {Can}, \citenamefont {Binicio\u{g}lu},\ and\
  \citenamefont {Shumovsky}}]{Klyachko:2008PRL}%
  \BibitemOpen
  \bibfield  {author} {\bibinfo {author} {\bibfnamefont {A.~A.}\ \bibnamefont
  {Klyachko}}, \bibinfo {author} {\bibfnamefont {M.~A.}\ \bibnamefont {Can}},
  \bibinfo {author} {\bibfnamefont {S.}~\bibnamefont {Binicio\u{g}lu}},\ and\
  \bibinfo {author} {\bibfnamefont {A.~S.}\ \bibnamefont {Shumovsky}},\ }\href
  {https://doi.org/10.1103/PhysRevLett.101.020403} {\bibfield  {journal}
  {\bibinfo  {journal} {Phys. Rev. Lett.}\ }\textbf {\bibinfo {volume} {101}},\
  \bibinfo {pages} {020403} (\bibinfo {year} {2008})}\BibitemShut {NoStop}%
\bibitem [{\citenamefont {Cirel'son}(1993)}]{Tsirelson:1993HJS}%
  \BibitemOpen
  \bibfield  {author} {\bibinfo {author} {\bibfnamefont {B.~S.}\ \bibnamefont
  {Cirel'son}},\ }\href {https://www.tau.ac.il/~tsirel/download/hadron.pdf}
  {\bibfield  {journal} {\bibinfo  {journal} {Hadron. J. Suppl.}\ }\textbf
  {\bibinfo {volume} {8}},\ \bibinfo {pages} {329} (\bibinfo {year}
  {1993})}\BibitemShut {NoStop}%
\bibitem [{\citenamefont {Abramsky}\ and\ \citenamefont
  {Hardy}(2012)}]{Abramsky:2012PRA}%
  \BibitemOpen
  \bibfield  {author} {\bibinfo {author} {\bibfnamefont {S.}~\bibnamefont
  {Abramsky}}\ and\ \bibinfo {author} {\bibfnamefont {L.}~\bibnamefont
  {Hardy}},\ }\href {https://doi.org/10.1103/PhysRevA.85.062114} {\bibfield
  {journal} {\bibinfo  {journal} {Phys. Rev. A}\ }\textbf {\bibinfo {volume}
  {85}},\ \bibinfo {pages} {062114} (\bibinfo {year} {2012})}\BibitemShut
  {NoStop}%
\bibitem [{Note2()}]{Note2}%
  \BibitemOpen
  \bibinfo {note} {An ideal (or sharp) measurement of an observable is one
  which does not disturb any jointly measurable observable (thus it is
  ``minimally disturbing'' \cite {Chiribella:2016IC}). An ideal observable is
  one in which all coarse-grainings can be ideally measured. It is important to
  stress that, independently of how difficult may it be to implement ideal
  measurements in practice \cite {Guryanova:2020Q} (and, apparently, we can do
  it pretty well \cite {Pokorny:PRL20,WangSAdv2022}), the concept of ideal
  observable is necessary in any maximally predictive physical theory allowing
  joint measurements}\BibitemShut {NoStop}%
\bibitem [{\citenamefont {Zeilinger}(2005)}]{Zeilinger:2005Nat}%
  \BibitemOpen
  \bibfield  {author} {\bibinfo {author} {\bibfnamefont {A.}~\bibnamefont
  {Zeilinger}},\ }\href {https://doi.org/10.1038/438743a} {\bibfield  {journal}
  {\bibinfo  {journal} {Nature}\ }\textbf {\bibinfo {volume} {438}},\ \bibinfo
  {pages} {743} (\bibinfo {year} {2005})}\BibitemShut {NoStop}%
\bibitem [{Note3()}]{Note3}%
  \BibitemOpen
  \bibinfo {note} {A measurement scenario is characterized by a number of
  observables, the cardinality of their respective sets of possible outcomes,
  and the description of which observables can be jointly measured. For
  example, the scenario with the smallest number of observables in which there
  can be KS contextuality (and Bell nonlocality) is the one with four
  dichotomic observables, $A,B,a,b$, such that the following pairs are jointly
  measurable: $\{A,B\}$, $\{A,b\}$, $\{a,B\}$, $\{a,b\}$.}\BibitemShut {Stop}%
\bibitem [{\citenamefont {Goh}\ \emph {et~al.}(2018)\citenamefont {Goh},
  \citenamefont {Kaniewski}, \citenamefont {Wolfe}, \citenamefont {V\'ertesi},
  \citenamefont {Wu}, \citenamefont {Cai}, \citenamefont {Liang},\ and\
  \citenamefont {Scarani}}]{Goh:2018PRA}%
  \BibitemOpen
  \bibfield  {author} {\bibinfo {author} {\bibfnamefont {K.~T.}\ \bibnamefont
  {Goh}}, \bibinfo {author} {\bibfnamefont {J.}~\bibnamefont {Kaniewski}},
  \bibinfo {author} {\bibfnamefont {E.}~\bibnamefont {Wolfe}}, \bibinfo
  {author} {\bibfnamefont {T.}~\bibnamefont {V\'ertesi}}, \bibinfo {author}
  {\bibfnamefont {X.}~\bibnamefont {Wu}}, \bibinfo {author} {\bibfnamefont
  {Y.}~\bibnamefont {Cai}}, \bibinfo {author} {\bibfnamefont {Y.-C.}\
  \bibnamefont {Liang}},\ and\ \bibinfo {author} {\bibfnamefont
  {V.}~\bibnamefont {Scarani}},\ }\href
  {https://doi.org/10.1103/PhysRevA.97.022104} {\bibfield  {journal} {\bibinfo
  {journal} {Phys. Rev. A}\ }\textbf {\bibinfo {volume} {97}},\ \bibinfo
  {pages} {022104} (\bibinfo {year} {2018})}\BibitemShut {NoStop}%
\bibitem [{\citenamefont {Navascu{\'e}s}\ \emph {et~al.}(2007)\citenamefont
  {Navascu{\'e}s}, \citenamefont {Pironio},\ and\ \citenamefont
  {Ac{\'\i}n}}]{NPA_PRL}%
  \BibitemOpen
  \bibfield  {author} {\bibinfo {author} {\bibfnamefont {M.}~\bibnamefont
  {Navascu{\'e}s}}, \bibinfo {author} {\bibfnamefont {S.}~\bibnamefont
  {Pironio}},\ and\ \bibinfo {author} {\bibfnamefont {A.}~\bibnamefont
  {Ac{\'\i}n}},\ }\href {https://doi.org/10.1103/PhysRevLett.98.010401}
  {\bibfield  {journal} {\bibinfo  {journal} {Phys. Rev. Lett.}\ }\textbf
  {\bibinfo {volume} {98}},\ \bibinfo {pages} {010401} (\bibinfo {year}
  {2007})}\BibitemShut {NoStop}%
\bibitem [{\citenamefont {Hall}(2010)}]{Hall:2010PRL}%
  \BibitemOpen
  \bibfield  {author} {\bibinfo {author} {\bibfnamefont {M.~J.~W.}\
  \bibnamefont {Hall}},\ }\href
  {https://doi.org/10.1103/PhysRevLett.105.250404} {\bibfield  {journal}
  {\bibinfo  {journal} {Phys. Rev. Lett.}\ }\textbf {\bibinfo {volume} {105}},\
  \bibinfo {pages} {250404} (\bibinfo {year} {2010})}\BibitemShut {NoStop}%
\bibitem [{\citenamefont {Hall}(2011)}]{Hall:2011PRA}%
  \BibitemOpen
  \bibfield  {author} {\bibinfo {author} {\bibfnamefont {M.~J.~W.}\
  \bibnamefont {Hall}},\ }\href {https://doi.org/10.1103/PhysRevA.84.022102}
  {\bibfield  {journal} {\bibinfo  {journal} {Phys. Rev. A}\ }\textbf {\bibinfo
  {volume} {84}},\ \bibinfo {pages} {022102} (\bibinfo {year}
  {2011})}\BibitemShut {NoStop}%
\bibitem [{\citenamefont {Costa~de Beauregard}(1977)}]{Costa:1977NC}%
  \BibitemOpen
  \bibfield  {author} {\bibinfo {author} {\bibfnamefont {O.}~\bibnamefont
  {Costa~de Beauregard}},\ }\href {https://doi.org/10.1007/BF02906749}
  {\bibfield  {journal} {\bibinfo  {journal} {Il Nuovo Cimento B}\ }\textbf
  {\bibinfo {volume} {42}},\ \bibinfo {pages} {41} (\bibinfo {year}
  {1977})}\BibitemShut {NoStop}%
\bibitem [{\citenamefont {Cramer}(1980)}]{Cramer:1980PRD}%
  \BibitemOpen
  \bibfield  {author} {\bibinfo {author} {\bibfnamefont {J.~G.}\ \bibnamefont
  {Cramer}},\ }\href {https://doi.org/10.1103/PhysRevD.22.362} {\bibfield
  {journal} {\bibinfo  {journal} {Phys. Rev. D}\ }\textbf {\bibinfo {volume}
  {22}},\ \bibinfo {pages} {362} (\bibinfo {year} {1980})}\BibitemShut
  {NoStop}%
\bibitem [{\citenamefont {Wood}\ and\ \citenamefont
  {Spekkens}(2015)}]{Wood:2015NJP}%
  \BibitemOpen
  \bibfield  {author} {\bibinfo {author} {\bibfnamefont {C.~J.}\ \bibnamefont
  {Wood}}\ and\ \bibinfo {author} {\bibfnamefont {R.~W.}\ \bibnamefont
  {Spekkens}},\ }\href {https://doi.org/10.1088/1367-2630/17/3/033002}
  {\bibfield  {journal} {\bibinfo  {journal} {New J. Phys.}\ }\textbf {\bibinfo
  {volume} {17}},\ \bibinfo {pages} {033002} (\bibinfo {year}
  {2015})}\BibitemShut {NoStop}%
\bibitem [{\citenamefont {Maudlin}(1992)}]{Maudlin:1992}%
  \BibitemOpen
  \bibfield  {author} {\bibinfo {author} {\bibfnamefont {T.}~\bibnamefont
  {Maudlin}},\ }in\ \href
  {https://doi.org/10.1086/psaprocbienmeetp.1992.1.192771} {\emph {\bibinfo
  {booktitle} {PSA 1992: Proceedings of the Biennial Meeting of the Philosophy
  of Science Association. Volume One: Contributed Papers}}},\ \bibinfo {editor}
  {edited by\ \bibinfo {editor} {\bibfnamefont {D.}~\bibnamefont {Hull}},
  \bibinfo {editor} {\bibfnamefont {M.}~\bibnamefont {Forbes}},\ and\ \bibinfo
  {editor} {\bibfnamefont {K.}~\bibnamefont {Okruhlik}}}\ (\bibinfo
  {publisher} {University of Chicago Press},\ \bibinfo {address} {Chicago,
  IL},\ \bibinfo {year} {1992})\ pp.\ \bibinfo {pages} {404--417}\BibitemShut
  {NoStop}%
\bibitem [{\citenamefont {Wheeler}(1986)}]{Wheeler:1986XXX}%
  \BibitemOpen
  \bibfield  {author} {\bibinfo {author} {\bibfnamefont {J.~A.}\ \bibnamefont
  {Wheeler}},\ }in\ \href {https://doi.org/10.1111/j.1749-6632.1986.tb12434.x}
  {\emph {\bibinfo {booktitle} {New Techniques and Ideas in Quantum Measurement
  Theory}}},\ \bibinfo {series} {Ann. N. Y. Acad. Sci.}, Vol.\ \bibinfo
  {volume} {480},\ \bibinfo {editor} {edited by\ \bibinfo {editor}
  {\bibfnamefont {D.~M.}\ \bibnamefont {Greenberger}}}\ (\bibinfo {address}
  {New York City, NY},\ \bibinfo {year} {1986})\ pp.\ \bibinfo {pages}
  {304--316}\BibitemShut {NoStop}%
\bibitem [{\citenamefont {Cabello}\ \emph {et~al.}(2016)\citenamefont
  {Cabello}, \citenamefont {Gu}, \citenamefont {G{\"u}hne}, \citenamefont
  {Larsson},\ and\ \citenamefont {Wiesner}}]{Cabello_Th_PRA2016}%
  \BibitemOpen
  \bibfield  {author} {\bibinfo {author} {\bibfnamefont {A.}~\bibnamefont
  {Cabello}}, \bibinfo {author} {\bibfnamefont {M.}~\bibnamefont {Gu}},
  \bibinfo {author} {\bibfnamefont {O.}~\bibnamefont {G{\"u}hne}}, \bibinfo
  {author} {\bibfnamefont {J.-{\AA}.}\ \bibnamefont {Larsson}},\ and\ \bibinfo
  {author} {\bibfnamefont {K.}~\bibnamefont {Wiesner}},\ }\href
  {https://doi.org/10.1103/PhysRevA.94.052127} {\bibfield  {journal} {\bibinfo
  {journal} {Phys. Rev. A}\ }\textbf {\bibinfo {volume} {94}},\ \bibinfo
  {pages} {052127} (\bibinfo {year} {2016})}\BibitemShut {NoStop}%
\bibitem [{\citenamefont {Chiribella}\ and\ \citenamefont
  {Yuan}(2016)}]{Chiribella:2016IC}%
  \BibitemOpen
  \bibfield  {author} {\bibinfo {author} {\bibfnamefont {G.}~\bibnamefont
  {Chiribella}}\ and\ \bibinfo {author} {\bibfnamefont {X.}~\bibnamefont
  {Yuan}},\ }\href {https://doi.org/10.1016/j.ic.2016.02.006} {\bibfield
  {journal} {\bibinfo  {journal} {Inf. Comput.}\ }\textbf {\bibinfo {volume}
  {250}},\ \bibinfo {pages} {15} (\bibinfo {year} {2016})}\BibitemShut
  {NoStop}%
\bibitem [{\citenamefont {Guryanova}\ \emph {et~al.}(2020)\citenamefont
  {Guryanova}, \citenamefont {Friis},\ and\ \citenamefont
  {Huber}}]{Guryanova:2020Q}%
  \BibitemOpen
  \bibfield  {author} {\bibinfo {author} {\bibfnamefont {Y.}~\bibnamefont
  {Guryanova}}, \bibinfo {author} {\bibfnamefont {N.}~\bibnamefont {Friis}},\
  and\ \bibinfo {author} {\bibfnamefont {M.}~\bibnamefont {Huber}},\ }\href
  {https://doi.org/10.22331/q-2020-01-13-222} {\bibfield  {journal} {\bibinfo
  {journal} {Quantum}\ }\textbf {\bibinfo {volume} {4}},\ \bibinfo {pages}
  {222} (\bibinfo {year} {2020})}\BibitemShut {NoStop}%
\bibitem [{\citenamefont {Pokorny}\ \emph {et~al.}(2020)\citenamefont
  {Pokorny}, \citenamefont {Zhang}, \citenamefont {Higgins}, \citenamefont
  {Cabello}, \citenamefont {Kleinmann},\ and\ \citenamefont
  {Hennrich}}]{Pokorny:PRL20}%
  \BibitemOpen
  \bibfield  {author} {\bibinfo {author} {\bibfnamefont {F.}~\bibnamefont
  {Pokorny}}, \bibinfo {author} {\bibfnamefont {C.}~\bibnamefont {Zhang}},
  \bibinfo {author} {\bibfnamefont {G.}~\bibnamefont {Higgins}}, \bibinfo
  {author} {\bibfnamefont {A.}~\bibnamefont {Cabello}}, \bibinfo {author}
  {\bibfnamefont {M.}~\bibnamefont {Kleinmann}},\ and\ \bibinfo {author}
  {\bibfnamefont {M.}~\bibnamefont {Hennrich}},\ }\href
  {https://doi.org/10.1103/PhysRevLett.124.080401} {\bibfield  {journal}
  {\bibinfo  {journal} {Phys. Rev. Lett.}\ }\textbf {\bibinfo {volume} {124}},\
  \bibinfo {pages} {080401} (\bibinfo {year} {2020})}\BibitemShut {NoStop}%
\bibitem [{\citenamefont {Wang}\ \emph {et~al.}(2022)\citenamefont {Wang},
  \citenamefont {Zhang}, \citenamefont {Luan}, \citenamefont {Um},
  \citenamefont {Wang}, \citenamefont {Qiao}, \citenamefont {Xie},
  \citenamefont {Zhang}, \citenamefont {Cabello},\ and\ \citenamefont
  {Kim}}]{WangSAdv2022}%
  \BibitemOpen
  \bibfield  {author} {\bibinfo {author} {\bibfnamefont {P.}~\bibnamefont
  {Wang}}, \bibinfo {author} {\bibfnamefont {J.}~\bibnamefont {Zhang}},
  \bibinfo {author} {\bibfnamefont {C.-Y.}\ \bibnamefont {Luan}}, \bibinfo
  {author} {\bibfnamefont {M.}~\bibnamefont {Um}}, \bibinfo {author}
  {\bibfnamefont {Y.}~\bibnamefont {Wang}}, \bibinfo {author} {\bibfnamefont
  {M.}~\bibnamefont {Qiao}}, \bibinfo {author} {\bibfnamefont {T.}~\bibnamefont
  {Xie}}, \bibinfo {author} {\bibfnamefont {J.-N.}\ \bibnamefont {Zhang}},
  \bibinfo {author} {\bibfnamefont {A.}~\bibnamefont {Cabello}},\ and\ \bibinfo
  {author} {\bibfnamefont {K.}~\bibnamefont {Kim}},\ }\href
  {https://doi.org/10.1126/sciadv.abk1660} {\bibfield  {journal} {\bibinfo
  {journal} {Sci. Adv.}\ }\textbf {\bibinfo {volume} {8}},\ \bibinfo {pages}
  {eabk1660} (\bibinfo {year} {2022})}\BibitemShut {NoStop}%
\end{thebibliography}%


\end{document}